\documentclass[default,iicol]{sn-jnl}% Default with double column layout

\usepackage{natbib}
\usepackage[official]{eurosym}
\setcitestyle{authoryear,open={(},close={)},aysep={},citesep={;}}

\jyear{2021}%

\theoremstyle{thmstyleone}%
\theoremstyle{thmstyletwo}%

\theoremstyle{thmstylethree}%

\newcommand{\makeblue}[1]{\textcolor[rgb]{0,0,0}{#1}}

\begin{document}

\title[Article Title]{In-situ measurements of void fractions and bubble size distributions in bubble curtains.}


\author*[1]{\fnm{Simon} \sur{Beelen}}\email{s.j.beelen@utwente.nl}
\author[2]{\fnm{Martijn} \sur{van Rijsbergen}}

\author[2]{\fnm{Milo\v s} \sur{Birvalski}}

\author[2]{\fnm{Fedde} \sur{Bloemhof}}
\author*[1]{\fnm{Dominik} \sur{Krug}}\email{d.j.krug@utwente.nl}

\affil*[1]{\orgdiv{Physics of Fluids group}, \orgname{University of Twente}, \orgaddress{\street{Drienerlolaan 5}, \city{Enschede}, \postcode{7522NB}, \country{Netherlands}}}

\affil[2]{\orgname{Maritime Research Institute Netherlands}, \orgaddress{\street{Haagsteeg 2}, \city{Wageningen}, \postcode{6708 PM}, \country{Netherlands}}}


\abstract{
We report the development of a novel measurement system designed to measure bubble properties in bubble curtains (i.e. planar bubble plumes) \makeblue{in-situ alongside} acoustical measurements. Our approach is based on electrical, contact-based needle sensors in combination with an optical system. The latter is used for calibration and validation purposes. Correcting for the insensitive distance of the needle tips yields very good agreement between the two approaches in terms of the local void fraction and bubble size distributions. Finally, the system is employed to study bubble plumes evolving from three different hose types. All hoses display consistent self-similar behaviour with spreading rates increasing with increasing gas flow. The spreading is further found to be significantly higher when the bubble plumes originated from a porous hose compared to the two other hose types featuring either discrete holes or nozzle elements.}




\maketitle

\section{Introduction}\label{sec1}
Bubble curtains are widely employed across diverse engineering applications such as controlling the movement of fish \citep{noatch2012non}, limiting the spreading of floating debris \citep{spaargaren2018bubble} or liquid spills \citep{lo1997effect}, or to mitigate  salt water intrusions \citep{abraham1962reduction}. Besides these, the use of bubble curtains to reduce underwater noise pollution is of particular interest, especially with the recent push to increase  offshore windfarm capacity. The pile driving during the installation of such windfarms generates significant noise emissions. If unmitigated, these can harm marine mammals within a radius of 100 m and still cause significant  disturbance at distances up to 50 km around a pile driving site \citep{bailey2010assessing,duarte2021soundscape}. The use of bubble curtains to reduce the noise impact during construction is a proven concept already that is also frequently employed in practice \citep{OSPAR,wursig2000development,tsouvalas2020underwater}. However, the operation of bubble curtains is  expensive and costs can easily surpass \euro{}100.000 per pile depending on parameters such as the depth of the water and local currents \citep{strietman2018measures}. There is hence significant interest to achieve performance improvements to reduce these expenses during construction. Another related aspect is the need for validated models to ensure that the pile-driving operation reliably complies with the relevant regulations, e.g. regarding the maximum permitted sound pressure level in order to protect marine life 

The interaction with the sound critically depends on parameters of the bubbles such as their size, their distribution and the resulting variations in void fraction \citep{commander1989linear}. These parameters, in turn, can vary drastically depending on local 
conditions, such as water depth and current, but importantly may also change in response to variations in the water quality \citep[e.g.][]{zhang2014novel}, e.g. the concentrations of dissolved gas, surfactants and salt \citep[]{winkel2004bubble,craig1993effect,firouzi2015quantitative}. Within the bubble curtain these parameters can vary due to for example coalescence and break up \citep{camarasa1999influence} and due to the decrease in hydrostatic pressure the bubbles experience when rising \citep{zheng2010local}. Experimental investigations aimed at a better understanding of the relevant physics of the sound-bubble interaction therefore ideally combine both acoustic and hydrodynamic measurements simultaneously in the same setup and location. 
The practical challenge in doing so lies in the fact that the emitted noise of pile driving typically peaks between $100$--$500\, \text{Hz}$ \citep[e.g.][]{bailey2010assessing,matuschek2009measurements}. As a consequence, the relevant wave lengths are in the order of 10--3m, which necessitates the use of large basins to avoid spurious confinement effects on the acoustics. 
Given these difficulties, the acoustic insertion loss of bubble curtains is commonly measured outdoors in lakes or the sea, where a detailed characterisation of the bubble properties is either lacking entirely \citep[]{dahne2017bubble,wursig2000development,stein2015hydro,lucke2011use} or complemented from a separate laboratory test \citep[e.g.][]{rustemeier2012underwater}.
In an effort to push beyond these limitations, this present paper presents the development, calibration and validation of a novel measurement setup to characterize bubble curtains with good accuracy and at large scale up to 1.8m in width. Crucially, the new device can be operated \textit{in situ} to characterize the bubble curtain alongside acoustic tests.

There exists a multitude of different approaches to measure bubble properties (see the overviews by \citet{boyer2002measuring} and \citet{mudde2010advanced}). At the most general level, these can be classified into intrusive and non-intrusive measurement techniques.
Non-intrusive measurement techniques based on image analysis are typically used in configurations where the camera can be placed outside the flow domain \citep[e.g.][]{ferreira2012statistical,besagni2016bubble,wang2018behavior} and require sophisticated algorithms to deal with overlapping bubbles \citep[e.g.][]{lau2013development,zou2021recognition,de2018efficiency}. Severe limitations in terms of void fraction and/or depth of field remain however, precluding an underwater camera outside of the bubble curtain as a feasible option for large bubble curtains. Image analysis can be used semi-intrusively and locally if a small camera unit is placed within the bubble curtain.
Intrusive methods such as optical fibre probes \citep[e.g.][]{luther2004bubble,pjontek2014bubble,magaud2001experimental,enrique2005accuracy}, hot film anometry \citep[e.g.][]{rensen2005hot,wang2001measurement} and electrical probes \citep[e.g.][]{steinemann1984application,munoz2017development,huang2018local,revankar1993theory,tompkins2018wire,tyagi2017experimental} are generally capable of handling higher void fractions. In particular, electrical probes have been employed in bubble curtains before, e.g. by \citet{chmelnizkij2016schlussbericht}  who performed measurements in a basin with a diameter of 5 $\mathrm{m}$ and a depth of 4.8 $\mathrm{m}$. A total of 16 probes were spaced across the width of the bubble curtain at distances of 20--240 $\mathrm{mm}$ depending on the height above the nozzles. At 14 out of 16 sensor locations, a second sensor was placed at a short vertical distance (6.4 $\mathrm{mm}$) to provide estimates on bubble rise velocities and sizes via the contact times at individual sensors and the delay of the bubble hits between them. 
Such a two sensor arrangement is attractive and simple in principle. However, in practice a calibration is required to account for interactions between bubble and sensor. And even then interpretation at the level of a single bubble remains very difficult since the measured chord length will vary even for the same bubble depending on if it is pierced closer to its edge or its center. 
\citet{besagni2016estimation} proposed a clever calibration method for dual optical fibre probes that focuses on determining the statistics of the bubble size distribution from the measured chord length distribution. They use image analysis to establish a relation between the aspect ratio of bubbles as a function of their equivalent diameter, which serves as an input to the calibration. The procedure then consists of updating an estimated (log-normal) bubble size distribution iteratively until the predicted chord length distribution matches the experimentally obtained one. 
In a similar fashion, other methods reported in literature such as the maximum entropy method have been developed to transform the chord length distribution to a bubble size distribution \citep[e.g.][]{tyagi2017experimental,santana2006characteristic}.

In the following, we will describe the hardware aspects of our measurement device and the test rig in section \ref{sec2} and provide details on the methods for calibration and data analysis in section \ref{sec3}. This will be followed by validation and initial measurement results in section \ref{sec4} and finally our conclusions (section \ref{sec5}). 

\section{Experimental setup}\label{sec2}

\subsection{General considerations}
As mentioned previously, typical void fractions (in the percent range) and lateral dimensions (on the order of meters) do not lend themselves to the application of optical techniques if the camera is to be placed outside the bubble plume. Immersing the optical system inside the bubbly flow region can mitigate some of the issues. However, doing so introduces disturbances to the flow and can only capture a small region of the flow per  camera unit.
In order to capture the bubble distribution across the full height and width of the bubble curtain, conductivity based electrical sensors are therefore the most appealing choice. Even though each individual sensor only represents a point measurement, their low cost and simple operation renders combining many of them into a larger array to achieve a high spatial resolution.  For our purpose, we opted to combine the electrical sensors with an underwater camera that simultaneously captures a part of the bubble field in the center of the curtain. Doing so has three main benefits: \textit{(i)} The images provide a way to independently verify the results from the electrical sensors -- seeing is believing. \textit{(ii)} The optical measurements allow us to obtain the input required for the sensor calibrations under exactly the same conditions. This  greatly improves the robustness and accuracy of the method.
\textit{(iii)} The camera recordings provide a means to gauge to what extent very small bubbles, which may not be picked up by the electrical sensors, are of relevance in a particular configuration. This is an important aspect in interpreting the sensor results, that would otherwise be unaccounted for.

\subsection{Electrical probes}\label{subsec22}
\subsubsection{Probe design and operation}\label{subsubsec221}
\begin{figure}[!ht]
\centering
\includegraphics[width=\columnwidth]{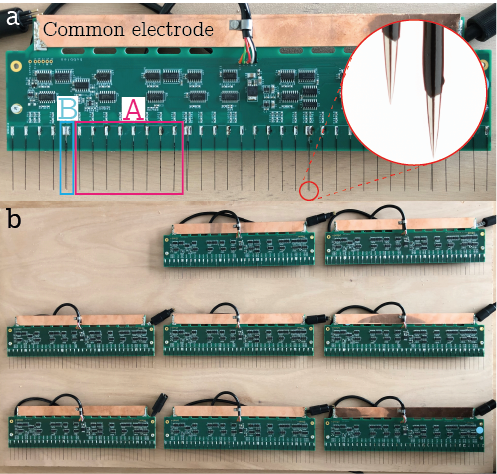}
\caption{a) One of the PCB's used for the measurements consisting of single needles (A) and double needles (B). b) Overview of multiple available PCB's}\label{fig:PCB}
\end{figure}

The electrical probes are made of coated stainless steel acupuncture needles. The acupuncture needles have a diameter of $D_n=0.12\, \mathrm{mm}$ and length $L_n=40\, \mathrm{mm}$. In order to insulate the needle shaft, a black coating is applied by dipping the needle in paint which also acts as a metal primer. The thickness of the coating ($\sim 20\,\mu \mathrm{m}$ ) is controlled by the speed at which the needle is lifted out of the bath. Finally, approximately 1 mm of the coating covering the tip of the needle is stripped to expose this part (see detail in Figure \ref{fig:PCB}a). 
The needle probes are then soldered onto a 300 mm wide printed circuit board (PCB). The PCB serves to keep the probes in place but also compactly hosts the required circuitry. Afterwards the PCB is fully coated in transparent epoxy to render it waterproof. Figure \ref{fig:PCB}a shows a single PCB, which holds a total of 40 needles spaced by 8.5 mm. Most locations contain only a single needle (labelled 'A' in figure \ref{fig:PCB}a), but every 10th position is equipped with a vertically staggered double needle ('B') to also measure the bubble velocity.

A potential (typically 3V) is applied between the needles and a common electrode in the water and each electrical probe is connected to a comparator circuit. Gas encapsulating the tip of the probe increases the resistance between the tip of the probe and the common electrode resulting in a switch of the output signal of the comparator. It is most straightforward to drive the circuits in direct current (DC) mode and this is used for the results presented here. We noticed, however, that this leads to the build up of tarnish on the tip of the probe, which significantly hampers operation and eventually results in an insulated tip. It is possible to restore functionality by cleaning between experiments and this has been done for the results reported here. Yet, a more elegant and sustainable solution is, to operate the circuits in alternating current (AC) mode. Doing so proved to effectively minimize the impact of tarnish on the measurement results. A diagram of a circuit suitable for both DC as well as for AC operation (with an alternating square wave) is shown in appendix \ref{secA1}. 

The complete system allows for simultaneous operation of up to 240 needles across 6 PCB's covering a range of 1.8 m in total (see Figure \ref{fig:PCB}b). However, in the measurements reported here only 3 PCB's were used. Each sensor is read out with 10 kHz in order to resolve a relevant timescale of about 1ms (based on a bubble velocity of 1 $\mathrm{ms^{-1}}$ and a size of 1 mm). To achieve the data transfer, the binary signal of each comparator circuit is serialized on the PCB, such that the outputs of 40 needles are sent consecutively to one common circuit above water, that gathers the data of 6 PCBs and combines the bits of each board into 40 bytes per sample.

\subsubsection{Characterisation of the probe tips}\label{subsubsec222} 
\begin{figure}[!ht]
\centering
\includegraphics[width=\columnwidth]{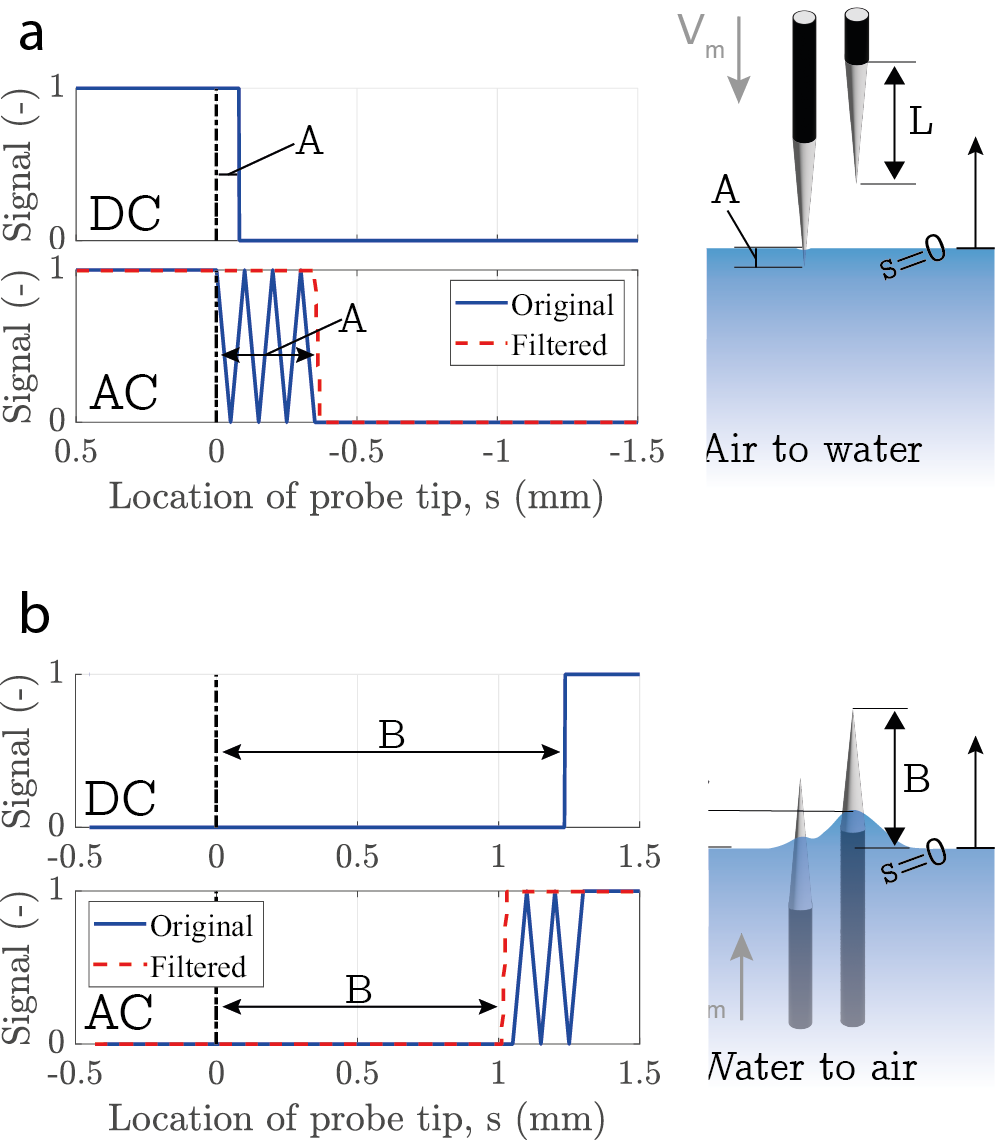}
\caption{\makeblue{DC and AC signal as measured by a needle probe, every 10001th sample has been shown for clarity. A schematic of the situation is shown alongside} for a) Air to water and b) water to air}\label{fig:Needlestuff2}
\end{figure}

The stripped part of the tip has a finite extent $L \approx 1\, \mathrm{mm}$ causing some uncertainty to the precise positions of the gas-water interface relative to the needle at which the signal will switch when going in and out of the bubble, respectively. In order to characterise this accurately, tests have been performed simulating the piercing process of a bubble by moving the needle through a water surface (see figure \ref{fig:Needlestuff2}). The needles were moved using a micro stage at a velocity of $V_m=50 \,\mathrm{\mu ms}^{-1}$ and the time at which the tip hit the surface was determined optically.

Results from these tests for both the DC actuation and the AC actuation are shown in Figure \ref{fig:Needlestuff2} along with the filtering of the AC signal. The probe signal is plotted vs the tip location $s$, with $s = 0\, \mathrm{mm}$ corresponding to the water surface. From these data, it becomes clear that
the probe detects water, corresponding to "0" signal, as soon as only a small part of the tip is exposed to water. Concretely, this is reflected in the `switching distances' $A$ from the probe tip being short for the air to water transition in Figure \ref{fig:Needlestuff2}a, whereas the counterpart $B$ for the reverse direction is comparable to $L$. In the ideal case, the switching position would be the same for both directions, i.e. $A=B$. We can define a `lag distance' based on the difference $\Delta = \vert A-B\vert$. A non-zero $\Delta$ implies that bubbles smaller than this value cannot be detected. Practically, this turns out to be of little relevance as from our experience such small bubbles tend to not get pierced in the first place. However, the lag also results in an underestimation of the bubble contact time and hence the void fraction, which needs to be accounted for. We found that the values of $\Delta$ from the present quasi-static tests were significantly affected by the menisci forming at the needle and results will likely not transfer to the dynamic bubble interaction with curved surfaces. We therefore decided to determine the precise value of $\Delta$ as part of a calibration procedure. Obviously, reducing the stripped length $L$ would also help mitigate this issue, but we found that decreasing the exposed surface area significantly increased the problems with tarnishing during operation.\\

\begin{figure*}[!htb]
\centering
\includegraphics[width=\textwidth]{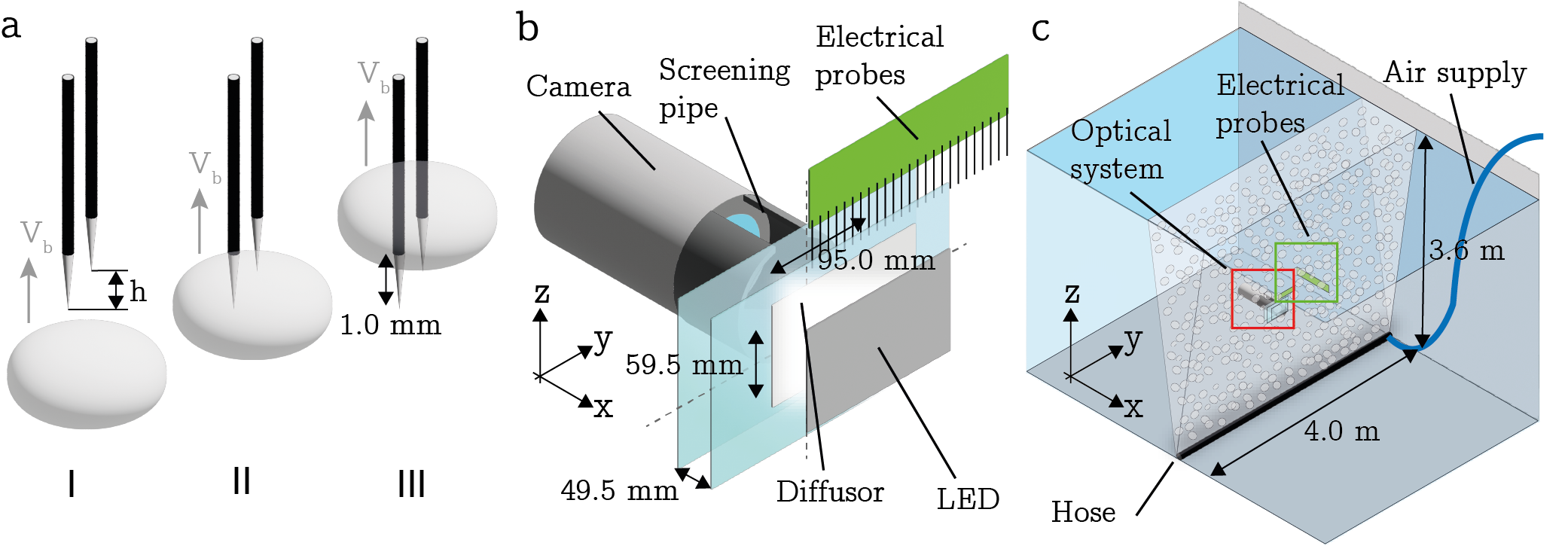}
\caption{a) Electrical probes, I a bubble approaches a set of double needles both needles are in full contact with water (signal 0 0), II Bubble encapsulates the tip of the longer needle (signal 1 0), III bubble left longer needle and encapsulates the shorter needle (signal 0 1). b) Optical system. c) Experimental setup with the bubble curtain hose, air supply, optical system and electrical probes}\label{fig:Setup}
\end{figure*}

The double probes (see Figure \ref{fig:PCB}a group B) are used to determine the bubble rise velocity and the chord length of the pierced bubble. The rise velocity can be found by $V_b=\frac{h}{t_2-t_1}$ where $h$ is the difference in height between the two probes (see Figure \ref{fig:Setup}a) and $t_1$ and $t_2$ are the arrival times of the bubble on the two needles. The chord length is determined by $C_b=V_bT_{con}$ where $T_{con}$ is the contact time of the bubble. For an accurate measurement $h$ is a vital parameter for which we  can obtain an experimental estimate by slowly moving the double probes out of the water (see figure \ref{fig:Needlestuff2}b). From the time difference in the signals and the known velocity $V_m$ we found $h = 0.85\, \mathrm{mm}$, which is in good agreement with values obtained from imaging the probe tips.

\subsection{Optical system}\label{subsec23}

The optical system is schemetically shown in Figure \ref{fig:Setup}b and consists of a Magma G-235 camera with a KOWA 16 mm lens in a watertight casing. The framerate of the camera is set to $\sim 30\,\mathrm{ Hz}$ and the resolution is $1920 \times 1200 \, \text{pix}^2$. The bubbles pass through the measurement volume which is enclosed by 2 transparent plexiglass plates leading to a measurement volume of $V_{meas}=95.0 \times 59.7 \times 49.5\,\mathrm{mm}^3$. The measurements have been calibrated using a transparent ruler which has been held against the front and back plate of the volume. 
Backlight illumination is provided through a custom made LED panel that is placed behind a diffusor plate. In order to keep the optical path free of bubbles, the region between the camera and the measurement volume is shielded by a screening pipe, which has a narrow opening at the top to let entrapped air escape. The optical system is combined with a single PCB located above the measurement volume spanning half the measurement volume (see Figure \ref{fig:Setup}b) to enable direct comparison of the results. \makeblue{Note that half of the length of this PCB extends beyond the plexiglas plates. This helped us ensure that the presence of the optical setup did not interfere with the measurement by confirming that the mean measured void fraction did not vary along this PCB.} The entire arrangement is then placed in the center of the plume such that the optical measurement domain aligns with the hose generating the bubble curtain as shown in Figure \ref{fig:Setup}c.

\subsection{Bubble curtain test facility}\label{subsec21}
Experiments with a bubble curtain have been carried out in the so-called Concept Basin (CB) at MARIN (MAritime Research Institute Netherlands). The CB measures LxWxD: 220x4x3.6 $\mathrm{m}^3$. A 4 m long aerator hose connected to the air supply was placed over the width of the basin on the bottom. In most experiments a `porous'-type aerator hose was used. Additional test were performed with a PVC pipe fitted with nozzles (Festo UC-1/8 silencers) every 100 mm and a conventional PVC pipe with holes of 1 mm in diameter drilled every 50 mm (see Appendix \ref{secB1} for images). Pressurised air was supplied by a compressor coupled to a Festo MS6-LFM-1 filter at normalised (to standard pressure) flow rates in the range of $0.55-1.67\, \mathrm{L m^{-1} s^{-1}}$, which were measured using a thermal mass flow sensor (Bronkhorst F-203AC). 
The entire measurement system was attached to a platform that could translate vertically enabling measurements between 0.2 and 3.6 m above the bottom of the basin. An overview of the entire setup is given in Figure \ref{fig:Setup}c. The figure also introduces the coordinate system used, where the height $z$ is measured from the top of the hose, $y$ runs along the hose and $x$ indicates the transversal direction of the bubble curtain.

\section{Post processing and analysis }\label{sec3}

\subsection{Electrical probes}\label{subsec31}

\subsubsection{Void fraction}\label{subsubsec311}
The output signal of an individual probe, $S_i$, is either 0 (water) or 1 (air). This means that the most straight forward estimate of the void fraction of probe $i$ is the average of the signal in time
\begin{equation}
    \overline{\alpha}^*_{n,i}= \frac{1}{T}\int_{T} S_{i}(t) dt,
    \label{eq:alfap}
\end{equation}
where $T$ is the averaging period. Subscript $n$ is used to indicate results from the needle measurements and the overline denotes time averaging throughout the paper.

Due to the finite lag distance $\Delta$, Eq. \ref{eq:alfap} will result in a slight underestimation of the actual void fraction. A correction can be made based on the number of bubbles $N_{T}$ that hit the probe during $T$, since for every pierced bubble a small part is not measured. Provided that $\Delta$ and the average bubble rise velocity $V_{r}$ are known, the total void fraction, $\bar\alpha_{n,i}$ is given by
\begin{equation}
    \overline{\alpha}_{n,i}=  \frac{1}{T} \left(\int_{T} S_{i}(t)dt
    +N_{T}\frac{\Delta}{V_{r}}\right). 
    \label{eq:alfai}
\end{equation}
which represents our measurement of the local void fraction at the location of probe $i$. \makeblue{The relevancy of the correction term is demonstrated based on our results in section \ref{sec:subsec41}}. As a caveat, it should be noted again that Eq. \ref{eq:alfai} does not include contributions of bubbles with chord lengths smaller than $\Delta$ as these will not be detected by the needle probes.

\subsubsection{Bubble size distribution}\label{subsubsec312}
\begin{figure*}[htb!]
\centering
\includegraphics[width=\textwidth]{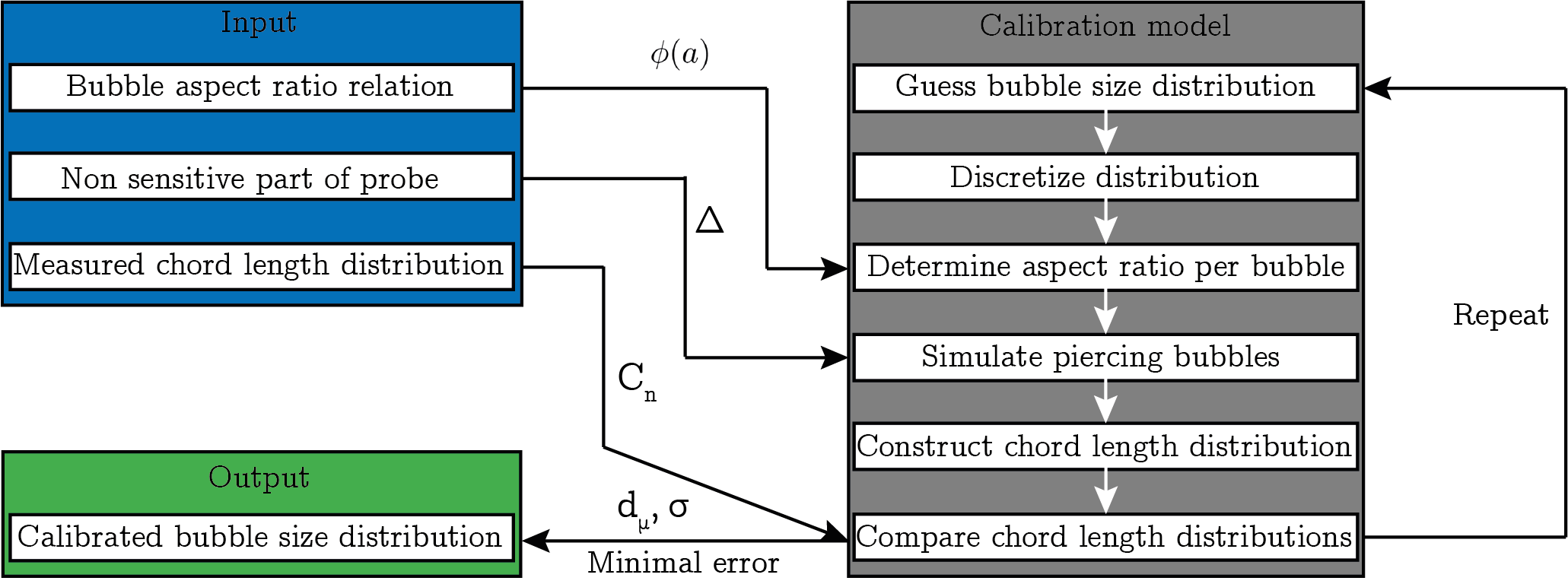}
\caption{Schematic of calibration algorithm }\label{fig:Calibration}
\end{figure*}

The probes measure a chord length distribution $C_{m}$ by using the rise velocity measured by both needles and the contact time of the bubble on the longer needle (generally the one that will be hit first). The chord length depends on the size of the bubbles, their shape but also on the location at which a particular bubble is pierced. The bubble size distribution is therefore not measured directly by the electrical probes but needs to be reconstructed based on $C_{m}$ and other parameters. In order to achieve this, we adopt a method outlined in \citet{besagni2016estimation}. 

The main steps of the procedure are outlined in Figure \ref{fig:Calibration}. Besides $C_{m}$, other required input parameters are an empirical relation for the typical bubble aspect ratio $\phi$ as a function of the bubble size, and the lag distance $\Delta$.  

The algorithm assumes that the distribution of the volume equivalent bubble diameter $d_{eq}$ can be approximated by a log-normal distribution. This choice is motivated by findings of \citet[]{mandal2005comparative}, who showed that a log-normal distribution is the most appropriate in flows where the bubble size is determined by break up and coalescence of bubbles, e.g. bubbles produced by nozzles in bubbly flows. 
The log-normal distribution is given by
\begin{equation}
    f=\frac{1}{d_{eq} \sigma \sqrt{2\pi}} \exp{\left(- \frac{(\ln{d_{eq}}-\ln{d_{\mu}})^2}{2\sigma^2}\right)}
    \label{eq:lognorm}
\end{equation}
with the \makeblue{shapefactor and the median of the distribution, $\sigma$ and $d_{\mu}$ respectively, being the free parameters}. The goal is then to determine the combination of these two parameters that is most consistent with the measured chord length distribution $C_m$. To this end,  $51\times 51$ variations of both variables are tested in the relevant range (as estimated by the results from the camera images) of $0.2<\sigma<0.3$ and $2\, mm<d_{\mu}<4\, mm$. For each combination, $N=100$ bubbles are considered with sizes $d_{eq}$ equally spaced between the lower measurement limit (0.96 mm) and 4 times the bubble radius where the log normal distribution has a maximum such that the $i$-th bubble appears with a probability
\begin{equation}
f_i=\frac{1}{d_{eq,N}-d_{eq,1}}\int_{d_{eq,i}}^{d_{eq,i+1}}f\textrm{d}d_{eq}.
\end{equation}

In the next step, for each of these bubbles a corresponding ellipsoidal shape is determined based on the aspect ratio $\phi \equiv b/a$ (with $a$,$b$ denoting the major and minor axes, respectively), which is given empirically in the form $\phi(a)$. We determine the ellipse properties based on equivalence of the cross sectional area resulting in \makeblue{$d_{eq} = \sqrt[3]{ab\sqrt{ab}}= \sqrt{ab}$}. Note that this differs from the choice in \citet{besagni2016estimation} who employed $d_{eq}=\sqrt[3]{a^2 b}$. The difference between these two approaches can be interpreted as a different choice for the out of plane axis, which is equal to $a$ in the case of \citet{besagni2016estimation}, while our definition implies that this dimension equals $\sqrt{ab}$. The main reason for our choice was that the large spread in the $\phi$ data appears inconsistent with assuming rotational symmetry around the minor axis for the bubbles (as implied by setting the out-of-plane dimension equal to $a$). 

Next, the vertical chord length is determined at 100 random locations for each bubble assuming a horizontal orientation of the major axis. Finally, the expected chord length distribution for that specific bubble $C_i$ is obtained after subtracting $\Delta$. The probability of hitting a bubble scales with its projected area $A_i = \frac{\pi}{4} \sqrt{a^3b}$.
The simulated chord length distribution $C_{sim}$ across all $N$ bubbles can therefore be constructed by summing the chord length contributions of the $N$ individual bubbles (${C_i}$) weighted by their projected area $A_i$ and by their probability $f_i$ according to
\begin{equation}
    {C_{sim}}=\frac{\sum_{i=1}^N {C_{i,j}}A_{i,j}f_{i,j}}{W\sum_{j=1}^{N_b}\sum_{i=1}^N {C_{i,j}}A_{i,j}f_{i,j}},
\end{equation}
where $N_b$ and $W$ respectively denote the number and width of the bins used for the chord length distributions. 
Finally the difference between the simulated and measured chord length distribution is quantified as the error
\begin{equation}
    E=\sum_{j=1}^{N_b} \vert{C_{sim,j}}-{C_{n,j}}\vert.
\end{equation}
and the combination of $\sigma$ and $d_{\mu}$ that minimizes $E$ is selected as the bubble size distribution $B_n$ corresponding to $C_n$. The entire procedure runs within minutes on a typical desktop computer, such that there was no need to go beyond the present brute-force approach. 

The two required input parameters $\Delta$ and $\phi(a)$ are determined with the use of the optical system. For the lag distance, $\Delta$, this was done by performing the above steps for varying values of the lag distance and comparing the output to a bubble distribution obtained from the camera images. We observed the best agreement for $\Delta = 0.96\, \mathrm{mm}$, which is also consistent with the considerations in section \ref{subsubsec222}. This value was therefore adopted throughout. 

The aspect ratio relation is determined by analyzing the camera images of the non-overlapping bubbles for every measurement. A moving average over the aspect ratio data of 1000 bubbles is then  fitted with a function of a similar form to that proposed by \citet{besagni2016estimation}. Details on this fit and corresponding results will be presented in section \ref{sec:ARfit}.

\subsection{Optical system}\label{subsec32}

\subsubsection{Image processing}\label{subsubsec321}
\begin{figure*}[!htb]
\centering
\includegraphics[width=\textwidth]{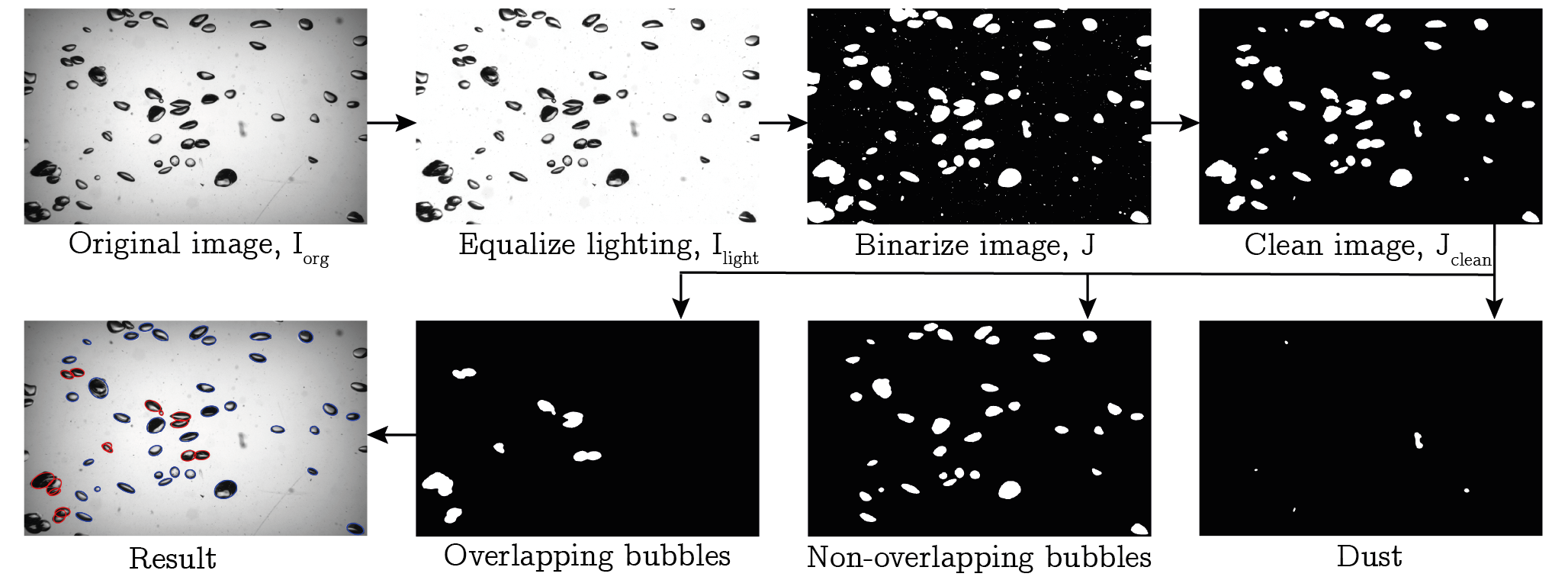}
\caption{Graphical overview of the used image analysis algorithm. In the final image, detected overlapping bubbles are indicated in red and the non overlapping ones in blue} \label{fig:Imagealgorithm}
\end{figure*}

The camera images are analysed in order to extract the void fraction, bubble size distribution and aspect ratio. The procedure used in doing so is outlined graphically in Figure \ref{fig:Imagealgorithm}. 

Starting from the original \makeblue{grayscale} image $I_{org}$, the lighting is equalized using a background image $I_{bg}$ according to 

\begin{equation}
I_{light}=I_{org} \frac{\text{max} (I_{bg})}{I_{bg}}.
\end{equation}

The equalised image $I_{light}$ is then binarized by comparing the local pixel value to $\text{max}(I_{bg})$ \makeblue{giving the binary image $J$ with $J =1$ indicating the presence of a bubble}. At this point,  small spots ($d_{eq}\leq 1\, \mathrm{mm}$) are removed from the image since these almost exclusively correspond to small dirt particles in the water. This is confirmed by visual inspection and by comparison to recordings made without bubbles present in the flow.
Consequently, bubbles touching the outer edge of the image are eliminated \makeblue{by setting $J =0$ at the corresponding locations} to yield the cleaned up image $J_{clean}$ (see Figure \ref{fig:Imagealgorithm}). %, \makeblue{i.e. the areas in the binary image which are 1 and touching the edge are replaced by 0}. 
In order to ensure that this step does not affect the void fraction estimate \makeblue{(removing the bubbles at the edge would lead to an underestimation of the void fraction)}, we compensate for the removed bubbles by adjusting the effective measurement volume per image by

\makeblue{
\begin{equation}
V_{adj}=V_{meas}\frac{\sum_J{J}_{clean}}{\sum_J J},
%V_{adj}=V_{meas}\frac{\langle{J}_{clean}\rangle}{\langle{J}_{bin}\rangle},
\label{eq:vadj}
\end{equation}}
where the sum $\sum_J$ is over the entire binarized image. $V_{meas}$ is the total measurement volume and $V_{adj}$ denotes the \makeblue{adjusted reference} volume used for determining the void fraction of the processed image. \makeblue{Note that Eq. \ref{eq:vadj} is based on the assumption that the average void fraction is uniform across the imaging region.} %\makeblue{$V_{adj}$ disregards the volume which is masked by the bubbles at the edge.} , $\langle{J}_{clean}\rangle$ is the average value of the \makeblue{total} binarized clean image and $\langle{J}_{bin}\rangle$ is the average value of the \makeblue{total} binarized image after removing the described areas. 
The clean image will be split up into `dust', `non-overlapping bubbles' and `overlapping bubbles'. The dust can be identified readily since the lack of reflection leads to lower gray values in this case. The non-overlapping bubbles are recognized by the solidity $\mathcal{S}$ of the patches, i.e. the ratio of the patch surface area to that of its convex hull. The criterion for non-overlapping bubbles is $\mathcal{S}>0.97$ and an ellipse with equal surface area is fitted to these cases. The remaining patches are then considered as clusters of two or more overlapping bubbles and their treatment is detailed in the following.

\subsubsection{Cluster deconstruction}

The goal of the cluster deconstruction is to describe the overlapping bubbles as a group of overlapping ellipses. In practice, there is a limit to what extent this is possible. In particular, the method presented here only identifies bubbles that form part of the outer contour of the cluster such that larger clusters of many bubbles can not be dealt with appropriately.

\begin{figure*}[!htb]
\centering
\includegraphics[width=\textwidth]{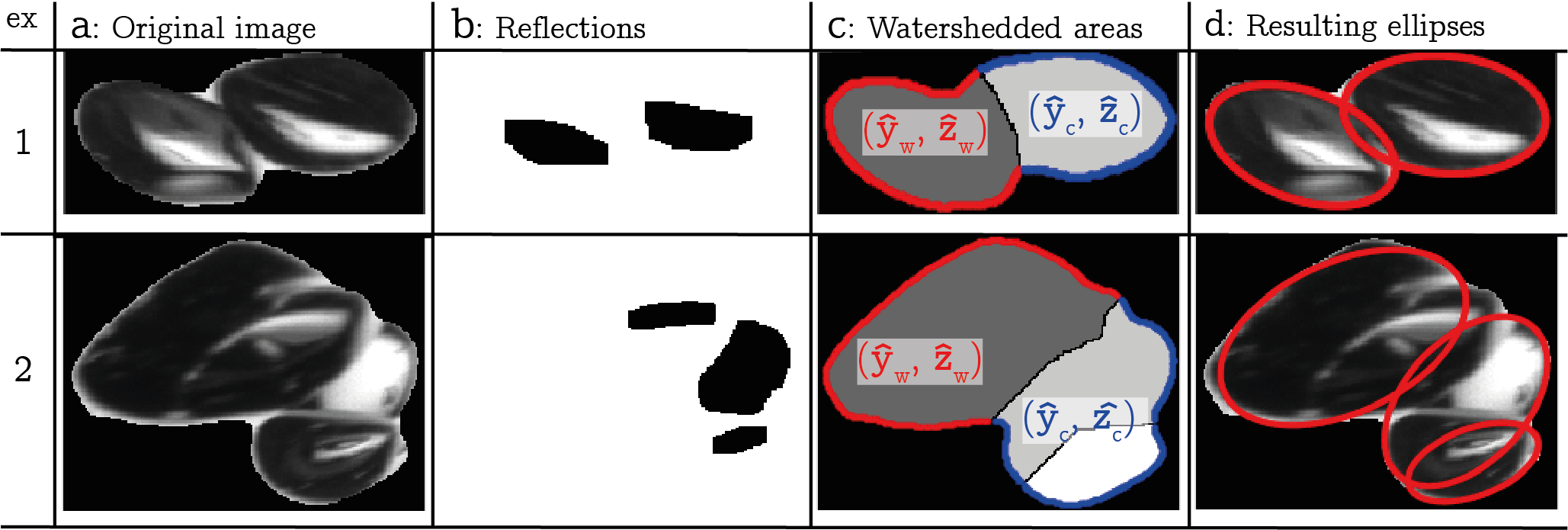}
\caption{Illustration of the bubble cluster deconstruction algorithm. 2 examples have been given: example 1 shows an almost perfect fit, example 2 shows a less successful application of the algorithm. a) The original images. b) Recognition of the bubble reflections. c) Watershedded areas constructed using the reflections. The coordinates of the contours are shown. d) The resulting ellipses }\label{fig:Watershed}
\end{figure*}

To illustrate our method, two examples are given in Figure \ref{fig:Watershed}. We make use of the fact, that the bubble images are not uniform but feature a brighter reflection spot. We identify these spots via simple thresholding (Figure \ref{fig:Watershed}b) and then use them as starting points (minima) for the watershedding technique \citep{meyer1994topographic}, which is widely employed in this context \citep{lau2013development, karn2015integrative}. The watershedding technique fills the contour of the cluster pixel by pixel starting from the outline of the reflections. Once adjacent watershedded areas touch, the areas will no longer grow in the direction in which they touched leading to a segmented cluster.% The segmented cluster is now roughly divided into it's components, the contour is thus divided into components as well.

In Figure \ref{fig:Watershed}c the parts of the deconstructed sample clusters are shown. We denote the coordinates of the segment of the contour pertaining to a single watershedded region as $(\bold{\hat{y}_w},\bold{\hat{z}_w})$ and use $(\bold{\hat{y}_c},\bold{\hat{z}_c})$ to refer to the remainder of the contour of the cluster, the hat denotes the use of the local image coordinate system. The task at hand is then to fit an appropriate ellipse to $(\bold{\hat{y}_w},\bold{\hat{z}_w})$. A general ellipse with arbitrary orientation and position is defined by $c_1 y^2+ c_2 z^2 +c_3 yz +c_4y+c_5 z -1=0$, where $c_1, \dots, c_5$ denote the 5 independent parameters. The most straightforward way to fit an ellipse is therefore to minimize the residual $\vert\vert \bold{P}_w\vert\vert_2$ of the system 
\begin{equation}
    \bold{c}\bold{M}_w-\bold{1}=\bold{P}_w,
    \label{eq:ellfit}
\end{equation}
where $\bold{M}_w=[\bold{{\hat{y}_w}}\circ\bold{{\hat{y}_w}}, \,\bold{{\hat{z}_w}}\circ\bold{{\hat{z}_w}}, \,\bold{{\hat{y}_w}}\circ\bold{{\hat{z}_w}}, \,\bold{{\hat{y}_w}}, \,\bold{{\hat{z}_w}}]$ and $\bold{c}=[c_1,\,c_2,\,c_3,\,c_4,\,c_5]$. 
However, this simple method does not penalize ellipses exceeding the contour of the cluster and therefore often leads to unphysical results.  This issue can be resolved noting that the values of $\boldsymbol{P}_w$ are negative if the fitted ellipse lies outside the watershedded contour. Ellipses exceeding the contour can therefore be penalized by putting a higher weight on negative values of the residual ($\boldsymbol{P}_w^-$) compared to their positive counterparts ($\boldsymbol{P}_w^+$). To also restrict ellipses from exceeding the contour at other regions of the cluster, we additional consider $\boldsymbol{P}_c$ defined analogous to equation \ref{eq:ellfit}. The full residual used in fitting the ellipses is therefore given by
\begin{equation}
    \vert\vert \boldsymbol{P}^*\vert \vert_2=a_1\vert\vert\bold{P_{w}^-}\vert\vert_2+ a_2 \vert\vert\bold{P_{w}^+}\vert\vert_2+a_3\vert\vert \bold{P_{c}^-}\vert\vert_2,
\end{equation}
where we used the weights $a_1=40$, $a_2=1$ and $a_3=30$. Note that $a_2>0$ is required to optimize to the contour shape, but since this does not apply for $\boldsymbol{P}_c^+$ it can be omitted. The ellipses fitted to the sample images with this method are shown in Figure \ref{fig:Watershed}d.

\subsubsection{Camera-based void fraction}\label{subsubsec322}
The void fraction averaged across the $i$-th image is determined by summing the volume of the $N(i)$ individual bubbles contained in it and dividing it by the volume associated to that image, such that
\begin{equation}
    \alpha_{c}(i)=\frac{\sum_{j=1}^{N(i)} V_{bub}^i(j)}{V^i_{adj}},
\end{equation}
with subscript `$c$' denoting quantities obtained from the camera images.
Note that in line with the discussion in section \ref{subsubsec312}, the bubble volume 
$V_{bub}=\frac{4}{3}\pi \left(\frac{d_{eq}}{2}\right)^3$ is again based on  $d_{eq}=\sqrt{ab}$ with the axes $a,b$ determined by the ellipse fits.

\section{Results and validation}\label{sec4}
Here, we present sample measurements to cross-validate camera and needle results and to illustrate the capabilities of the new system. To enable a meaningful comparison between camera and needle results, we consider a case with a moderate void fraction. This ensures that uncertainties in the image analysis due to overlaps remain limited while the number of bubbles is high enough to consider the mixture locally homogeneous. This trade-off was met best for a case with an air flow rate of $0.55\, \text{Lm}^{-1}\text{s}^{-1}$ at a height of $z = 3.4\, \text{m}$, which is therefore used as the test case in the following, unless specified otherwise.

\subsection{Void fraction}\label{sec:subsec41}
\begin{figure*}[!ht]
\centering
\includegraphics[width=\textwidth]{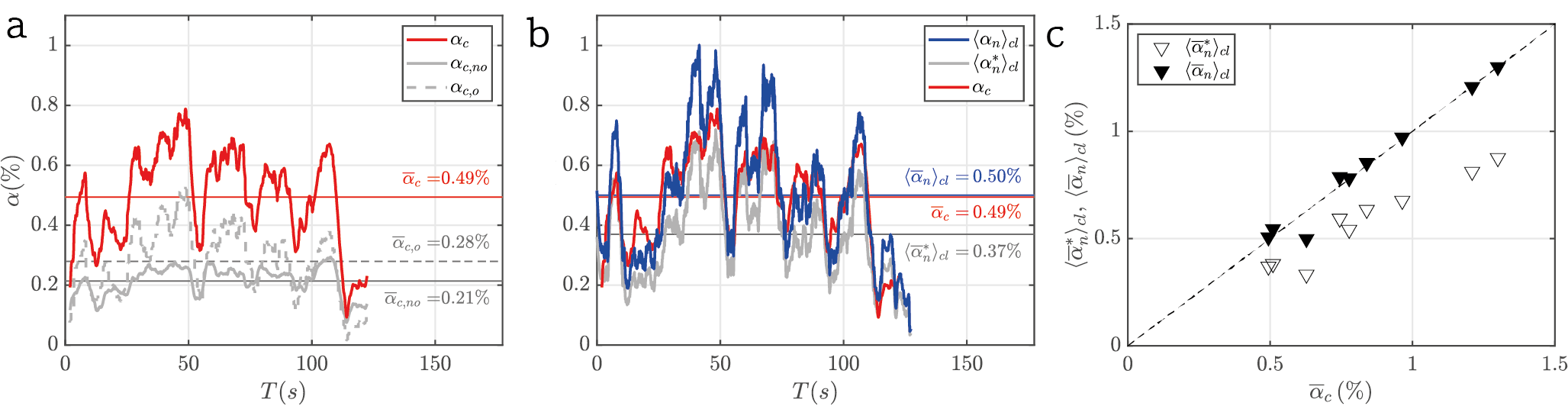}
\caption{a) Contribution of individual bubbles and bubbles in clusters to the void measured by the camera. b) Comparison of void fraction measured by the camera and by the electrical probes. \makeblue{c) Average uncorrected and corrected needle probe void fraction compared to the camera void fraction; the dashed line indicates the diagonal.}} \label{fig:Compcamneedles}
\end{figure*}

The void fraction is the most basic quantity extracted from the measurements. Results measured via the camera are shown in Figure \ref{fig:Compcamneedles}a as a function of time. We also show the contributions of non-overlapping bubbles ($\alpha_{c,no}$) and of overlapping bubbles ($\alpha_{c,o}$) individually. Even at the modest overall void fraction $\overline{\alpha}_c=0.49\,\%$, clusters are seen to contribute more than half of the total void fraction underlining the challenges in the optical approach. \makeblue{The increasing overlap for increasing void fraction is well known as illustrated for spherical objects by \citet{murai2001three}}. For reference, we note that assuming rotational symmetry around the minor axis to reconstruct the out-of plane dimension following \citet{besagni2016estimation} results in a higher estimate of 
$\overline{\alpha}_c  = 0.63\,\%$.

An appropriate reference to the camera results for the void fraction is the spatial average $\langle \alpha_n \rangle_{cl}$ over the 7 needle sensors positioned along the centerline of the bubble curtain directly above the camera measurement volume \makeblue{(see Figure \ref{fig:Setup}b)}.
A direct comparison between $\alpha_{c}$ (shown in red) and $\langle \alpha_n \rangle_{cl}$ (blue) is presented in Figure \ref{fig:Compcamneedles}b. Even though the measurement regions \makeblue{significantly differ in size and} do not fully overlap for the two methods, there is a very good agreement between the two methods. Some minor discrepancies are seen to occur whenever  $\langle \alpha_n \rangle_{cl}$ is high. These are likely a consequence of the difficulties in disentangling bubble clusters properly at these instances and we therefore expect the needle measurement to be more reliable in these conditions. The temporal mean $\langle \overline{\alpha}_n \rangle_{cl} = 0.50\%$ matches the camera result closely which validates the needle measurements. In particular, we note that with  $\langle \overline{\alpha}_n^* \rangle_{cl} = 0.37\%$ (see also grey line in Figure \ref{fig:Compcamneedles}b) the needle based estimate for the void fraction would be significantly lower without accounting for the lag distance $\Delta$. Here we used $\Delta=0.96\,\text{mm}$ based on fitting the bubble size distribution (see Section \ref{subsubsec312}) and $V_r= 1\,\text{ms}^{-1}$. \makeblue{In order to establish the validity of the lag correction beyond a single case, we present void fraction data for heights $z>2\, m$ and all three flowrates for the porous hose setup in Figure \ref{fig:Compcamneedles}c. When plotting against $\alpha_c$ as the reference, the corrected data ($\langle \alpha_n\rangle_{cl}$) nicely lines up on the diagonal indicating a good match across the full range of void fractions considered. Whereas, the deviations from the diagonal are considerable and results systematically too low if the lag correction is not included ($\langle \alpha_n^*\rangle_{cl}$).}

\subsection{Bubble aspect ratio}
\label{sec:ARfit}
\begin{figure}[!ht]
\centering
\includegraphics[width=\columnwidth]{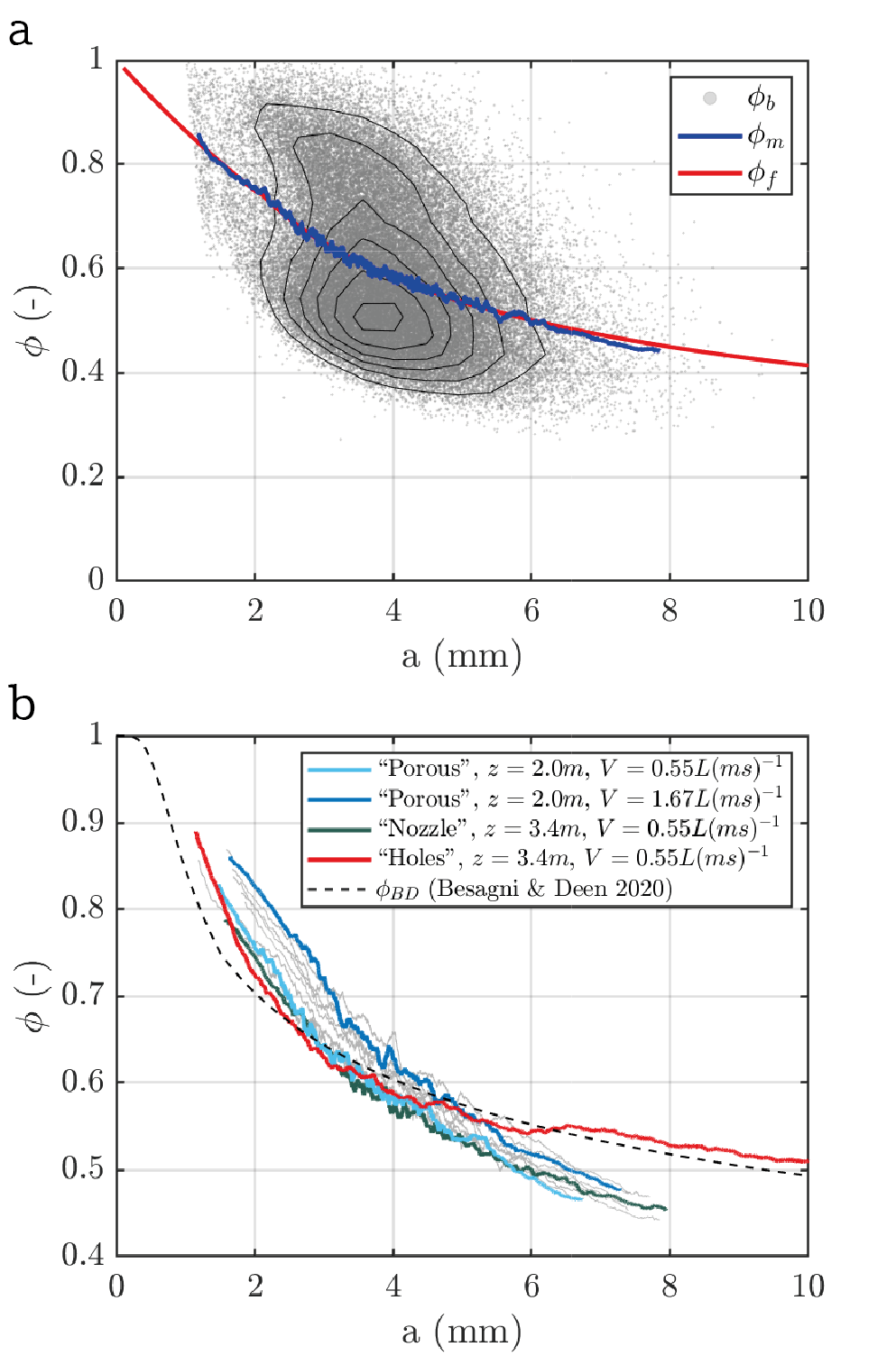}
\caption{a) Aspect ratio of individual bubbles with the contour of the joint pdf added to visualize the point density, the highest contour is at $1.2\,\mathrm{mm^{-1}}$ and steps between contour lines are $0.2\,\mathrm{mm^{-1}}$. The moving mean $\phi_m$ and the fit $\phi_f$ according to Eq. \ref{eq:Bes_and_Deen} are also shown. b) Variation of $\phi_m$ for different hoses, measurement heights $z$, and flow rates $V$. Grey lines represent additional case for the `porous' hose with $2.0\, \mathrm{m} \leq z \leq 3.4 \, \mathrm{m}$ and flow rates ranging within  $0.55\, \mathrm{Lm^{-1}s^{-1}} \leq V \leq  1.67 \,\mathrm{Lm^{-1}s^{-1}}$}\label{fig:Aspect}
\end{figure}

As described in section \ref{subsubsec312}, knowing the bubble aspect ratio as a function of the major axis ($\phi(a)$) is required to determine the bubble-size distribution. We can determine $\phi$ for individual bubbles from the camera images. Results for the test case (accumulated over a set of 1920 images) are shown in Figure \ref{fig:Aspect}a. Note that only results for non-overlapping bubbles are considered here as these data are more reliable. There is a considerable spread in $\phi(a)$ for individual bubbles and we therefore consider the moving mean $\phi_m$ over 1000 bubbles in order to reduce the scatter. This quantity can be fit using a composite expression similar to that used in  \citet{besagni2016estimation}:
\begin{equation}
\label{eq:Bes_and_Deen}
  \phi_f =
    \begin{cases}
      \kappa_1 a^2+\kappa_2 a + \kappa_3 & \text{if } a\leq \kappa_4\\
      \kappa_5 a^{\kappa_6} & \text{if } a>\kappa_4
    \end{cases}       
\end{equation}
Small bubbles with $a\to 0$ are not deformable and we therefore set $\kappa_3 =1$ while the remaining parameters are fit to the data. For the results shown in Figure 8a, we obtain $\kappa_1=0.0057\,\mathrm{mm^{-2}}$, $\kappa_2=-0.1378\,\mathrm{mm^{-1}}$, $\kappa_4=2\,\mathrm{mm}$, $\kappa_5=0.9810\,\mathrm{mm^{-\kappa_6}}$ and $\kappa_6=-0.3753$, which captures the variation in $\phi_m$ very well. 

We note, however, that the results for $\phi_m$ vary for different experimental conditions as shown in Figure \ref{fig:Aspect}b. 
For the case of the porous hose, there is a considerable dependence on the flow rate $V$, while the variation of $\phi_m$ at different heights is less pronounced. Additionally, also the details of the bubble generation are seen to affect the results.
Especially aspect ratio results for the `holes' hose differ from the two other configurations in particular for larger bubbles with $a \gtrsim 5\,\mathrm{mm}$. Similar effects have been noted in the literature before and in particular \citet{besagni2020aspect} proposed an expression to relate bubble aspect ratios to flow properties. Their prediction $\phi_{BD}$ requires the bubble Reynolds number and the E\"otv\"os number (relating surface tension and gravitational forces) as inputs. A result for $\phi_{BD}$ based on typical values of these quantities in our experiments is included in Figure \ref{fig:Aspect}b for reference. It agrees well  with the `holes' hose case, but does not represent the results for the other two configurations well. It is therefore highly beneficial to have the simultaneous optical measurements for every test case. This allows us to fit Eq. \ref{eq:Bes_and_Deen} for every configuration individually for the best accuracy.

\subsection{Bubble size distribution}
\begin{figure}[!ht]
\centering
\includegraphics[width=\columnwidth]{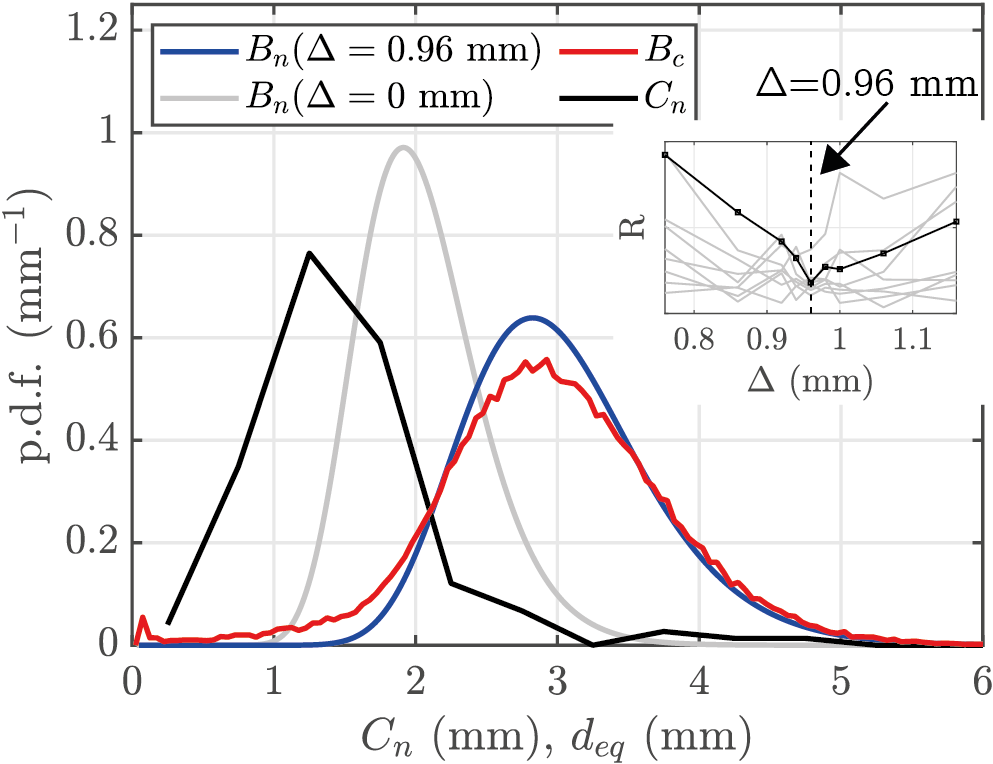}
\caption{Measured chord length distribution $C_n$ and the resulting bubble size distribution as measured using the needles $B_n$ compared to the bubble size distribution obtained via the camera $B_c$. Measured at z=3.4 m and V=0.55 Lm$^{-1}$s$^{-1}$}\label{fig:Input_calib}
\end{figure}

\begin{figure*}[!htb]
\centering
c\includegraphics[width=\textwidth]{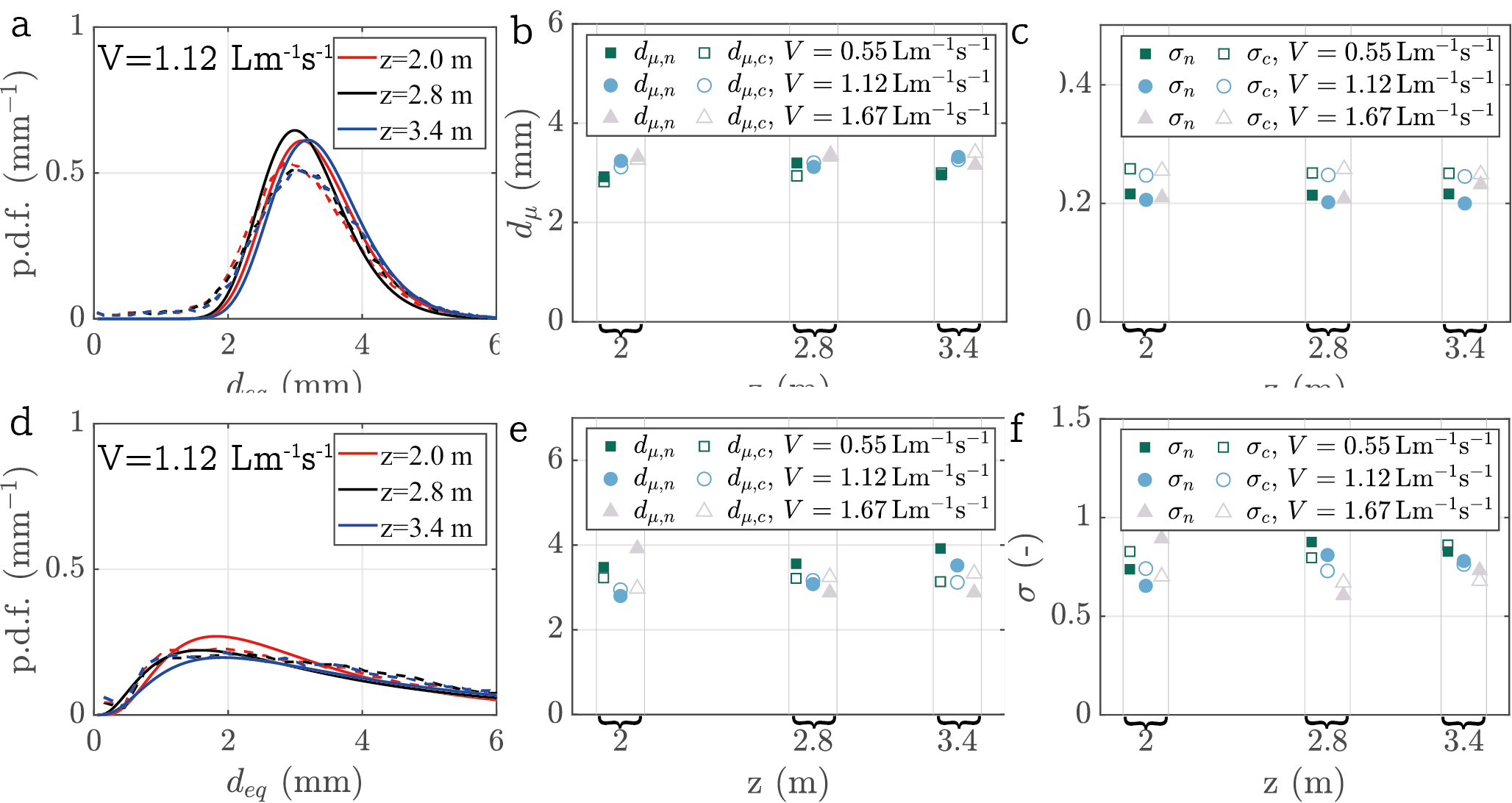}
\caption{Bubble size distributions obtained via the double probes compared to the camera images for the porous hose (a,b,c) and the holes hose (d,e,f). For every case $\phi_f(a)$ has been determined separately and $\Delta=0.96\, \mathrm{mm}$. a) Resulting bubble size distributions for a flowrate of $V=1.12\,\mathrm{Lm^{-1}s^{-1}}$, solid lines calibration result and dashed lines camera result b) $d_{\mu}$ for all considered cases compared to the camera. c) $\sigma$ for all considered cases compared to the camera d) Resulting bubble size distributions for a flowrate of $V=1.12\,\mathrm{Lm^{-1}s^{-1}}$. e) $d_{\mu}$ comparison. f) $\sigma$ comparison }\label{fig:Bubsizeneedlecam}
\end{figure*}

With $\phi(a)$ known, the procedure outlined in Section \ref{subsubsec312} can be employed to obtain a bubble size distribution based on the measured chord length distribution $C_n$. We can  cross-check these results vs. the bubble size distribution $B_c$, which is extracted from analysing the camera images. This comparison is presented in Figure \ref{fig:Input_calib} for the test case. A striking observation from this figure is by how much the distributions $C_n$ and $B_c$ differ, which underlines the need for appropriate postprocessing of the chord length data. How important it is to account for the lag distance $\Delta$ of the needle sensors becomes clear when considering $B_n(\Delta = 0\, \mathrm{mm})$, which represents the calibration result when ignoring this effect. The predicted size distribution in this case differs significantly from the reference $B_c$. To improve on this, we systematically vary $\Delta$ and monitor the residual (squared difference) $R$ between $B_n(\Delta)$ and $B_c$. This residual is plotted as the black line in the inset of Figure \ref{fig:Input_calib} and displays a minimum for a value of $\Delta=0.96\, \mathrm{mm}$, for which then also $B_n(\Delta=0.96\, \mathrm{mm})$ is found to be in very good agreement with $B_c$. Similar trends are also observed for other cases (shown as grey lines in the inset, see also Figure \ref{fig:Bubsizeneedlecam}), for which $R$ could be computed.  
Moreover, this value for $\Delta$ also matches the outcome of the probe tests in Section \ref{subsubsec222}. Additionally, we also found that $\overline{\alpha}_c$ and $\langle \overline{\alpha}_n \rangle_{cl}$ are in good agreement for this value and $\Delta=0.96\, \mathrm{mm}$ is therefore adopted for all measurements.

In Figure \ref{fig:Bubsizeneedlecam}, we present results on the bubble size distribution for different gas flow rates $V$ and heights $z$ for both the porous hose and for the holes hose. For the other cases, no comparison to camera results was possible, either due to the high void fraction (below $z=2\,\mathrm{m}$) or due to insufficient data (nozzle hose). The three samples for the holes hose at $V = 1.12\,\mathrm{Lm^{-1}s^{-1}}$ in Figure \ref{fig:Bubsizeneedlecam}a show similar agreement between $B_n$ and $B_c$ as observed in Figure \ref{fig:Input_calib}. A more systematic assessment is possible from Figures \ref{fig:Bubsizeneedlecam}b,c,e,f where the parameters of the log-normal distribution obtained from the calibration ($d_{\mu,n}$, $\sigma_{n}$) are compared to those resulting from fitting $B_c$ ($d_{\mu,c}$, $\sigma_{c}$). The agreement for $d_{\mu}$ (Figure \ref{fig:Bubsizeneedlecam}b) is very good and also the slight trend of increasing $d_{\mu}$ with increasing $V$ is captured faithfully for the most part. The correspondence is somewhat worse for $\sigma$ (Figure \ref{fig:Bubsizeneedlecam}c), where $\sigma_n$ is seen to underpredict $\sigma_c$ consistently by about 15-20\%. We believe that part of this discrepancy is due to the fact that very small bubbles ($d_{eq} \leq 1\,\mathrm{mm}$) are not picked up by the needle measurements. 
The bubble distributions originating from the holes hose (see samples in Figure \ref{fig:Bubsizeneedlecam}d) differ significantly from those of the porous hose, which was used for calibration.  Nevertheless, the agreement between camera and needle measurements overall remains equally as good for this case illustrating the robustness of our method. This can be judged by the results for $d_{\mu}$ (Figure \ref{fig:Bubsizeneedlecam}e) and $\sigma$ (Figure \ref{fig:Bubsizeneedlecam}f), where the former match somewhat less well compared to the porous hose, but the agreement for the latter is found to be better for the most part.

\subsection{Characterization of sample bubble plumes}
In the following, we present some representative results of plume characteristics based on the needle probes mounted on the two PCB's perpendicularly to the air-injection hose (see Figure \ref{fig:Setup}c).

\begin{figure*}[!ht]
\centering
\includegraphics[width=\textwidth]{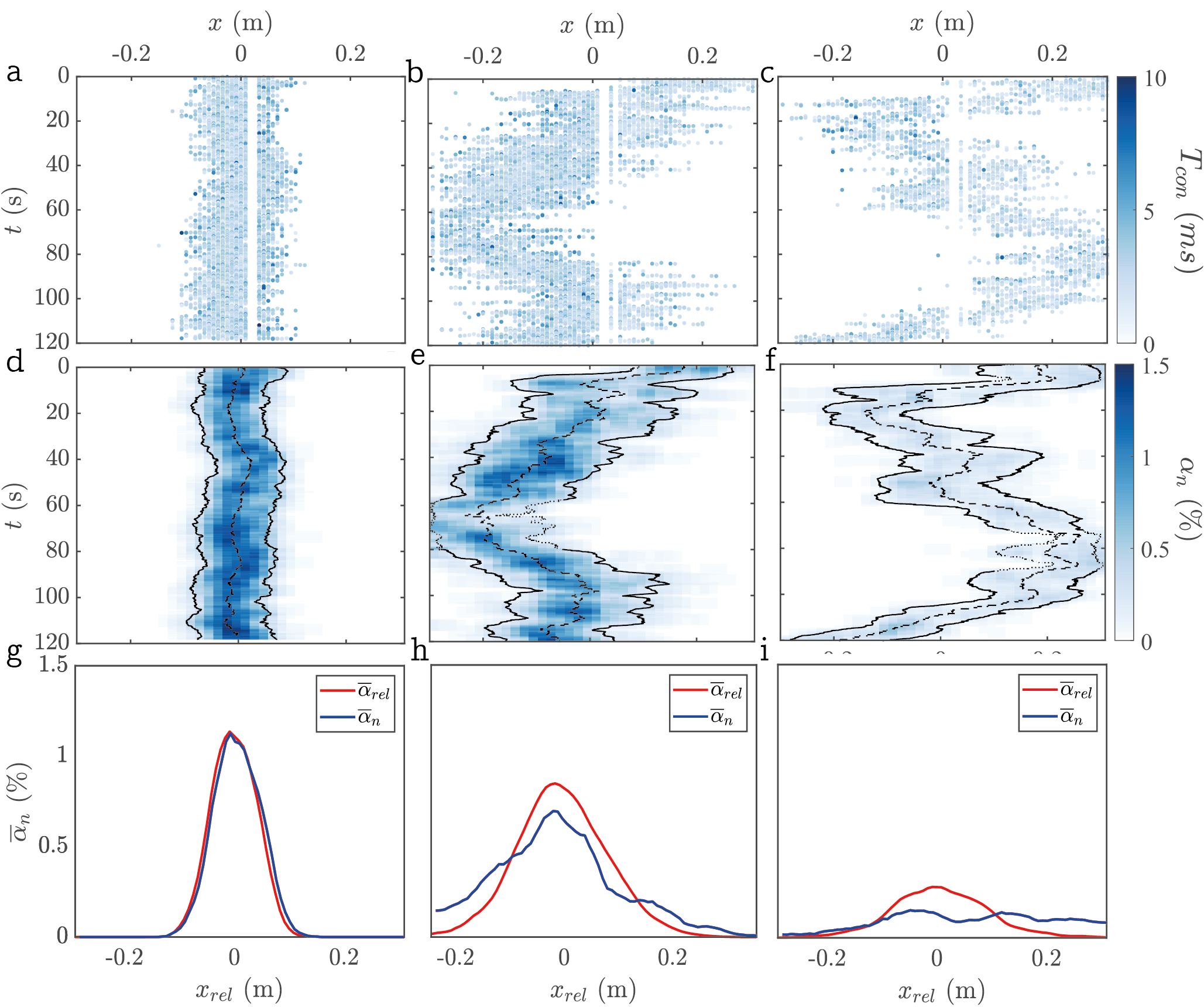}
\caption{Measurement results for three different measurement sets: at (a,d,g) $z=0.8\,\mathrm{m}$ and $V=1.12\,\mathrm{Lm^{-1}s^{-1}}$, (b,e,h) $z=2\,\mathrm{m}$ and $V=1.67\,\mathrm{Lm^{-1}s^{-1}}$, and (c,f,i) $z=2.8\,\mathrm{m}$ and $V=0.55\,\mathrm{Lm^{-1}s^{-1}}$.
(a,b,c) Spatiotemporal distribution of contact times $T_{con}$ at a needle (corrected for $\Delta$). (d,e,f) Local void fraction $\alpha_n$ as a function of time. (g,h,i) Time averaged void fraction based on conventional ($\alpha_{n}$ plotted vs. $x$) and conditioned ($\alpha_{rel}$ plotted vs. $x_{rel}$) averaging procedure. Note that the origin for $x_{rel}$ in (g,h,i) is chosen to align with the peak location of $\alpha_{n}$}\label{fig:Void_plates}
\end{figure*}

In Figure \ref{fig:Void_plates}, we chose three different measurement sets to illustrate differences in plume behaviour. The most basic quantity obtained from the needle sensors is the distribution of contact times $T_{con}$ in space and time. These are shown in Figure \ref{fig:Void_plates}(a-c) for the three cases. Already from these data, it is evident how the plume close to the hose in Figure \ref{fig:Void_plates}a is narrow and without significant lateral movement. At larger $z$ and $V$, the plume in Figure \ref{fig:Void_plates}b is somewhat wider but additionally also exhibits a slow waving motion in time. These fluctuations are most pronounced for the case shown in Figure \ref{fig:Void_plates}c, which is recorded even further from the hose and at a lower gas flow rate. Note that the `white stripes' in these figures are due to a fault of the needle probes at the corresponding locations. These data were recorded in DC mode and such probe failures were significantly reduced after switching to AC operation of the sensors.

In order to allow for a more quantitative evaluation, we determine the local void fraction $\alpha_n$ by averaging in time according to Eq. \ref{eq:alfai} with $T =5\,\mathrm{s}$ and taking the mean over 5 adjacent needle probes (corresponding to a distance of 30 mm). In doing so, we employ linear interpolation using the 2 adjacent needles to each side to fill in for the broken probes to avoid large gaps in the distribution of $\alpha_n$. The contours of $\alpha_n(x,t)$ corresponding to the data shown in Figure \ref{fig:Void_plates}a-c are displayed in Figure \ref{fig:Void_plates}d-f as a function of the measurement time. On this basis, it is possible to determine the location of the center of the bubble curtain ($x_{cen}$) from 
\begin{equation}
x_{cen}(t)=\frac{\int_{\Lambda}\alpha_n(x,t) x dx}{\int_{\Lambda} \alpha_n(x,t) dx},
\end{equation}
where the integration is over the width $\Lambda$ of the measurement rake ($600\,\mathrm{mm}$). The location of $x_{cen}(t)$ is indicated by the dashed lines in Figure \ref{fig:Void_plates}d-f.
Additionally, we define the `top-hat' scales for void fraction and plume width by 
\begin{equation}
\hat{\alpha}(t)=\frac{M(t)}{Q(t)}
\label{eq:alphahat}
\end{equation} 
and
\begin{equation}
\hat{w}(t)=\frac{Q(t)}{\hat{\alpha}(t)}
\label{eq:what}
\end{equation} 
based on the integrals $Q(t)=\int_{\Lambda} \alpha_n dx$ and $M(t)=\int_{\Lambda}\alpha_n^2dx$. Results for $\hat{w}$ are included in Figure \ref{fig:Void_plates}d-f as solid lines located at $ x_{cen}(t) \pm \hat{w}(t)/2$. We check for instances where the bubble curtain leaves the measurement region by verifying that $\vert x_{cen}(t) \pm \overline{\hat{w}}\vert \leq \Lambda/2$ and discard times where this does not hold from the processing (indicated by dotted outlines in Figure \ref{fig:Void_plates}d-f).

Finally, we present two types of temporal averages over the entire measurement time of 120 s in Figure \ref{fig:Void_plates}d-f). One is the conventional average in the laboratory frame of reference resulting in $\overline{\alpha}_n(x)$ as defined in Eq. \ref{eq:alfai}. In addition, we consider the conditioned average relative to the instantaneous centerline location $x_{cen}$, which we define as $\alpha_{rel}(x_{rel})$ where $x_{rel} = x - x_{cen}$ denotes the lateral coordinate relative to the centerline location. Naturally, the spatial distributions of $\overline{\alpha}_n$ and $\overline{\alpha}_{rel}$ agree closely in the first case (Figure \ref{fig:Void_plates}g), where there is very little movement of the centerline. Noticeable differences are visible for the case in Figure \ref{fig:Void_plates}h and in the most extreme case (Figure \ref{fig:Void_plates}i) the peak value of $\overline{\alpha}_{rel}$ is about twice that of $\overline{\alpha}_n$, which is significant despite the relatively poor statistical convergence in this case. These differences are of high relevance when trying to deduce acoustical properties of the bubble curtain as the sound interacts with the instantaneous distribution for which $\overline{\alpha}_{rel}$ is more representative. As pointed out by \citet{milgram1983mean} already, the slow movement of the centerline (or `wandering' as they called it) renders convergence of conventional averages poor and our measurements here are also certainly too short to fully characterise this wandering motion. \makeblue{From visual observations, it showed that the wandering motion was not uniform across the spanwise ($y$-direction). Especially in cases exhibiting stronger motion, the bubble curtain deformed into an S-shape across the basin.}

\begin{figure*}[!ht]
\centering
\includegraphics[width=0.8\textwidth]{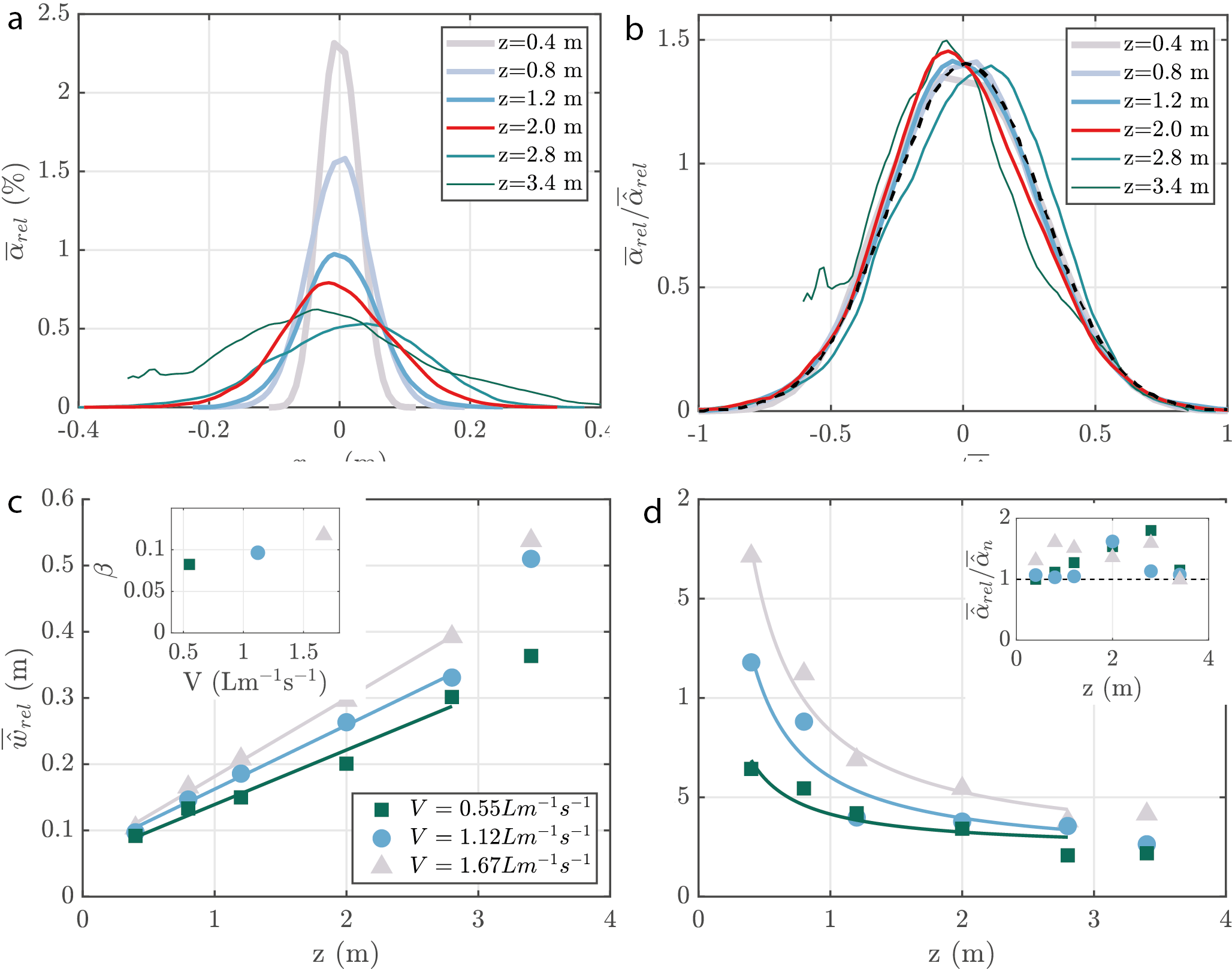}
\caption{a) Measured instantaneous void fraction at different heights for $V=1.67\,\mathrm{Lm^{-1}s^{-1}}$. b) Same data as in (a) rescaled by $\overline{\hat{w}}_{rel}$ and  $\overline{\hat{\alpha}}_{rel}$. The dashed black line is a fitted Gaussian which is characterized by the standard definition $\sigma_g=0.2845$ and the mean $\mu_g=0.01$. c) $\overline{\hat{w}}_{rel}$ as a function of $z$ for different air flow rates, markers show measured values and the solid lines show the best fit in the form of $\overline{\hat{w}}_{rel}=\beta z+\zeta$. d) $\overline{\hat{\alpha}}_{rel}$ as a function of $z$ for different air flow rates. The fit of the solid lines is in the form of $\overline{\hat{\alpha}}_{rel}=\upsilon z^{-1}+\xi$ } \label{fig:Void_height}
\end{figure*}

In Figure \ref{fig:Void_height}a, we present results for $\alpha_{rel}$ at different heights (with the red curve corresponding to the case shown in Figure \ref{fig:Void_plates}(b,e,h). Close to the hose, the peak void fraction exceeds 2\% and this value drops to about 0.5\% as the flow spreads upwards. The width of the plume at the highest measurement station ($z = 3.4\,\mathrm{m}$) exceeds the measurement domain and we also noticed that the presence of the free surface (located at $z = 3.6\,\mathrm{m}$) influenced the flow at this stage. The same data is re-plotted rescaled by $\hat{\alpha}_{rel}$ and $\hat{w}_{rel}$ (defined analogously to Eq. \ref{eq:alphahat} and Eq. \ref{eq:what}) in Figure \ref{fig:Void_height}b. Within the limits of the statistical convergence, there is a reasonable self-similar collapse for all data points except for the ones at $z = 3.4\,\mathrm{m}$. The distribution is well described by a Gaussian (dashed black line) and the same curve also fits the data at other values of $V$ and for other hose types very well (not shown).
In Figure \ref{fig:Void_height}c, time averaged values of the top-hat width as a function of the height for three different flow rates are shown. For all three cases, the data in the range  $0.4\,\mathrm{m} \leq z \leq 2.8\,\mathrm{m}$ can be well approximated by a linear fit of the form  $\overline{\hat{w}}=\beta z+\zeta$, which is consistent with the self-similar scaling for planar plumes \citep[i.e. neglecting the effect of slip velocity of the bubbles, see e.g. ][]{paillat2014entrainment}. Presumably due to interaction with the free surface, the last data point at $z = 3.4\,\mathrm{m}$ deviates slightly from the linear trend, in particular for the two higher gas flow rates. 

The spreading parameter $d\overline{\hat{w}}/dz=\beta$ (see inset of Figure \ref{fig:Void_height}c) is seen to increase with $V$. This trend as well as the magnitude of $\beta$ is consistent with results by \citet{cederwall1970analysis} for planar plumes based on velocity measurements. Similar findings (in terms of a Froude number dependence) are also reported for round plumes by \citet{kobus1968analysis} and \citet{milgram1983mean}. Finally, results for $\overline{\hat{\alpha}}_{rel}$ are shown in Figure \ref{fig:Void_height}d. The decay with $z$ is reasonably approximated by a $1/z$ dependence for these data, again consistent with the self-similar scaling for this quantity. The inset of Figure \ref{fig:Void_height}d shows the ratio between $\overline{\hat{\alpha}}_{rel}$ and $\overline{\hat{\alpha}}_{n}$. We expect this ratio to be large in cases with pronounced centerline movement. Results should be interpreted with care especially for the largest value of $V$ due to limited statistical convergence. It appears clear, however, that the relevance of centerline movement increases with increasing height, and presumably also with increasing $V$.

\begin{figure}[!ht]
\centering
\includegraphics[width=\columnwidth]{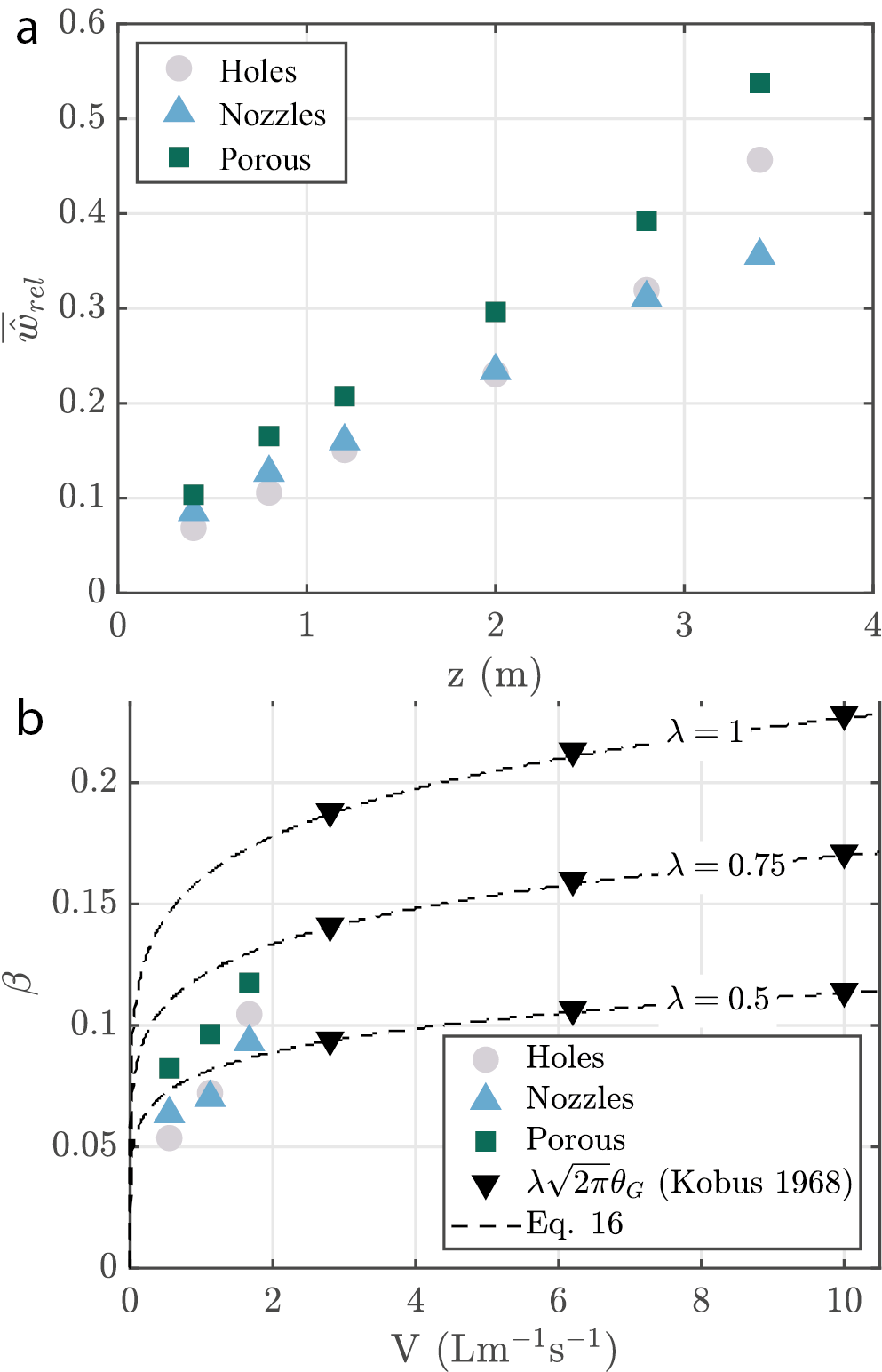}
\caption{ a) Top hat width of the three different hose types at $V=1.67\,\mathrm{Lm^{-1}s^{-1}}$. b) Spreading parameter for all hose types at varying gas flow rates \makeblue{ compared to the relation given in Eq. \ref{eq:entrain} and experimental data of \citet{kobus1968analysis}}} \label{fig:Beta}
\end{figure}

Finally, the spreading of the bubble plumes created with the three different hoses is compared in Figure \ref{fig:Beta}. Consistent with results at other flow rates (not shown), it can be seen from Figure \ref{fig:Beta}a that the plume originating from the porous hose is the widest at all heights, while differences remain small between the nozzle and the holes designs.
These trends are also reflected in Figure \ref{fig:Beta}b, where the growth rate of the bubble curtain width is shown for the three hose types. The growth rate of the holes hose and the hose with nozzles is very similar and significantly higher values of $\beta$ are observed for the porous hose. This difference does not appear to be related to differences in the bubble size distribution because there the largest deviations occur for the holes hose (see Figure \ref{fig:Bubsizeneedlecam}), while distributions for the other two cases are very similar.
A key difference between the configurations is that the porous hose is a continuous line source emitting air around its circumference, whereas the bubbles originate from discrete sources separated by respectively 50 mm and 100 mm for the holes and nozzle hoses. Even though these individual plumes merge within the first 1 m above the hose, especially the different plume evolution between the porous and nozzle hose suggest that these differences might affect the flow even at much larger heights. \makeblue{This is similar to \citet{wilkinson1979two} who also hypothesized a persisting influence of the initial buoyancy distribution on the plume evolution.}

\makeblue{The only known reference data on the spreading of planar bubble plumes known to us is by \citet{kobus1968analysis}. Their analysis is based on the velocity field yielding the velocity spreading parameter $\theta_G$, which is related to $\beta$ by $\beta=\lambda \sqrt{2\pi}\theta_G$. Here, the factor $\sqrt{2 \pi}$ accounts for the conversion from Gaussian variables (indicated by subscript `G') used in \citet{kobus1968analysis} to the top hat definition employed here. 
Additionally, the parameter $\lambda$ represents the ratio of the widths of the void fraction profile to that of the water velocity profile. Precise values of $\lambda$ are unknown with no measurements for planar plumes reported. Estimates used in the literature range from $\lambda=0.2$ (e.g in \citet{bohne2019modeling} based on \citet{ditmars1975analysis}), which seems unrealistically low as also pointed out in \citet{brevik2002flow} who suggest $\lambda=0.85$. The latter also corresponds more closely to the value of $\lambda=0.8$ reported for round plumes \citep{milgram1983mean}.
We compare our results for $\beta$ to those reported by \citet{kobus1968analysis} for three different values of $\lambda$ in Figure \ref{fig:Beta}b. Also included in the figure is the fit provided by \citet{kobus1968analysis}, which when expressed in terms of $\beta $ reads
\begin{equation}
\beta=\lambda \sqrt{2 \pi} 0.18 V^{0.15}.
\label{eq:entrain}
\end{equation}
A slightly modified version of this fit with an effective prefactor of 0.176 instead of 0.18 was given in \citet{brevik2002flow}, but the resulting difference is insignificant here in view of the uncertainty in $\lambda$.
With $3<V<10\,\mathrm{Lm^{-1}s^{-1}}$, the data of \citet{kobus1968analysis} falls into a different range compared to our results. From figure 13b, it does become clear, however, that the trends with respect to $V$ do not align very well between the two data sets. This is also reflected in the fact that the fit in Eq. \ref{eq:entrain} does not capture the present results well. The configuration in \citet{kobus1968analysis} resembles the 'holes' type hose, for which the fit is only of comparable magnitude if $\lambda = 0.5$, which appears unreasonably low given that the water flow is a direct result of the forcing via the rising bubbles.}

\section{Discussion and Conclusions}\label{sec5}
In summary, we report the development of a system capable of measuring the void fraction and the bubble size distribution in bubble curtains \emph{in-situ}. A key feature of our approach is the combination of a contact based sensor array with an optical system. This enables the  validation of the needle based measurements and allows us to calibrate the aspect ratio dependence $\phi(a)$ during the actual measurement. Our measurements revealed that accounting for the insensitive distance $\Delta$ of the needle tip was critical to obtaining accurate and reliable data from the electrical probes. When correcting for this effect, we found very good agreement between camera and needle based measurements of the local void fraction as well as of the bubble size distributions. We employ a statistical model that assumes a log-normal distribution for the latter, which describes the present data well. It appears possible, however, that modifications to this assumption might become necessary for other bubble generation methods and it is useful to monitor this via the optical measurements.

The system is then employed to measure the characteristics of bubble plumes originating from three different hose types. Our results show that especially at higher flowrates and further from the hose, it is critical to account for the meandering of the plume in order to obtain representative distributions as an input to an acoustic model. This can be achieved by conditioning on the instantaneous center. We find that for all three hose types, such conditioned void fraction distributions evolve in a self-similar manner up to the vicinity of the free surface. The spreading rates increase with increasing gas flow rate, which is consistent with previous reports on round \citep{milgram1983mean,kobus1968analysis,fraga2016influence} and planar bubble plumes \citep{kobus1968analysis}. However, we find that the spreading rate also depends on the method of bubble generation. In particular, the spreading parameter $\beta$ is substantially higher when the porous hose is employed. When interpreting these results, it should be kept in mind that here the analysis is based on the void fraction distribution while most other studies base the spreading on the velocity field. The widths of the velocity and of the void fraction profiles are not necessarily equal and their ratio can depend on details of the bubble generation as \citet{wu2021impact} have shown for single round plumes. \makeblue{ Quantitatively, we observe that the spreading rate is not well predicted by the relation of \citet{kobus1968analysis} in the range of air flow rates investigated here. Along with the differences for different hose types, this indicates that the gas flow rate alone is not sufficient to parameterize the spreading rate of bubble plumes.}

With the proposed measurement technique a data set with the hydrodynamical properties of different bubble curtains can be constructed. Future research should extend the use of this system to longer measurement times and larger arrays and focus on combining the measurements with acoustical measurements.

\backmatter

% \bmhead{Supplementary information}
% Not applicable

\bmhead{Acknowledgements}
The authors would like to thank Harm Jan Kamphof for his efforts in developing the optical system as part of his internship. And we would like to thank Christ de Jong for his careful read of the manuscript and his helpful comments.

\section*{Declarations}

\subsection*{Ethical Approval}
Not applicable
\subsection*{Competing interests}
The authors report no conflict of interest.
\subsection*{Authors' contributions}
Simon Beelen helped develop the measurement system, carried out the experiments, analysed and visualised the data and wrote the main manuscript.\\
Martijn van Rijsbergen contributed to the conceptual and detailed design of the measurement set-up and assisted with the experiments.\\
Milo\v s Birvalski helped develop the measurement system with a focus on the optical system and carried out the experiments.\\
Fedde Bloemhof helped develop the hardware of the measurement system and assisted with the experiments.\\
Dominik Krug: Conceptualization, Funding acquisition, Supervision, Writing original draft \& review and editing.\\
All authors reviewed the manuscript.

\subsection*{Funding}
This publication is part of the project AQUA (with project number P17-07) of the research programme perspectief which is (partly) financed by the Dutch Research Council (NWO).\\
This project is supported by the Netherlands Enterprise Agency (RVO) and TKI Wind op Zee.\\
 This project has received funding from the European Research Council (ERC) under the European Union's Horizon 2020 research and innovation programme (grant agreement No. 950111, BU-PACT).

\subsection{Availability of data and materials}
The data will be uploaded to arxiv.
\subsection*{Code availability}
The code is available upon request.

%\item Conflict of interest/Competing interests (check journal-specific guidelines for which heading to use)
%\item Ethics approval 
%\item Consent for publication
%\item Availability of data and materials
%\item Authors' contributions
%\end{itemize}

\begin{appendices}

\section{Circuitry electrical probes}\label{secA1}

\begin{figure}[!ht]
\centering
\includegraphics[width=\columnwidth]{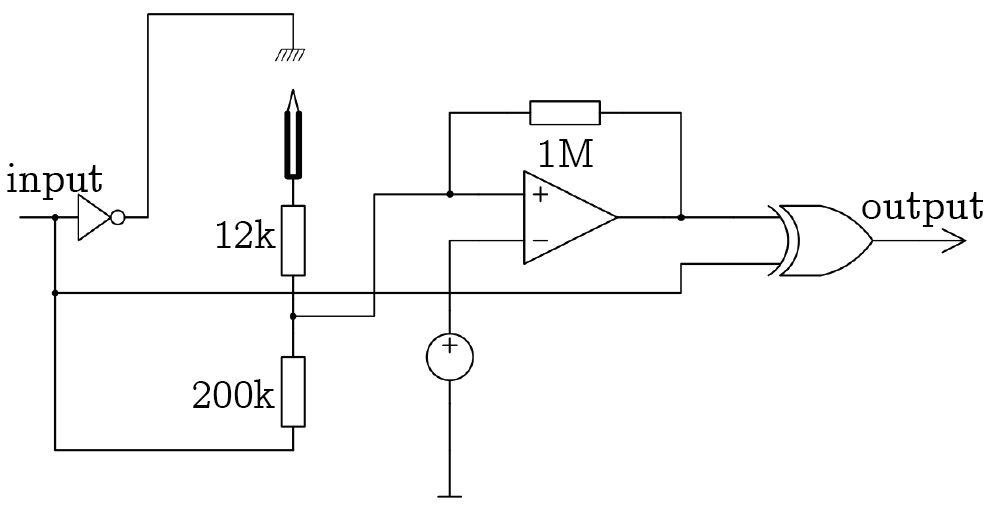}
\caption{The circuitry used by the needle probes, can be actuated either by a constant input(DC) or by an alternating input (AC)}\label{fig:Circuit}
\end{figure}

\section{Pictures hoses}\label{secB1}

\begin{figure}[!ht]
\centering
\includegraphics[width=\columnwidth]{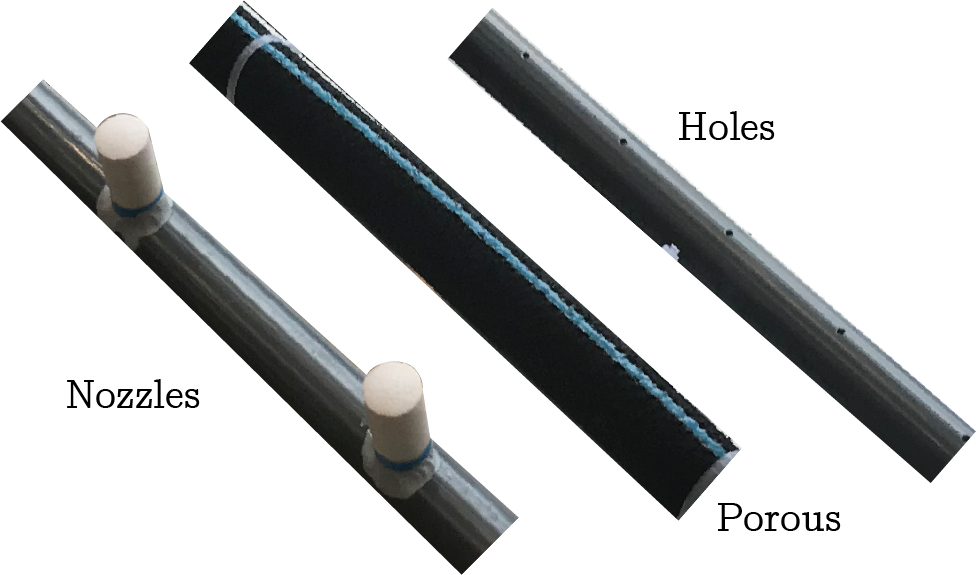}
\caption{The three different hoses used in the experiments}\label{fig:Circuit}
\end{figure}




\end{appendices}


\bibliography{sn-bibliography}% common bib file


\end{document}